\documentclass[conference]{IEEEtran}
\IEEEoverridecommandlockouts
\usepackage{cite}
\usepackage{amsmath,amssymb,amsfonts}
\usepackage{bbm}
\usepackage{algorithmic}
\usepackage{graphicx}
\usepackage{textcomp}
\usepackage{algorithmic}
\usepackage{algorithm}
\usepackage{multirow}
\usepackage{xcolor}
\DeclareMathOperator*{\argmin}{argmin}

\def\BibTeX{{\rm B\kern-.05em{\sc i\kern-.025em b}\kern-.08em
    T\kern-.1667em\lower.7ex\hbox{E}\kern-.125emX}}
\begin{document}

\title{Dataset Refinement for Improving the Generalization Ability of the EEG Decoding Model
\thanks{This research was supported by the Challengeable Future Defense Technology Research and Development Program through the Agency For Defense Development (ADD) funded by the Defense Acquisition Program Administration (DAPA) in 2024 (No.912911601) was partly supported by the Institute of Information \& Communications Technology Planning \& Evaluation (IITP) grant, funded by the Korea government (MSIT) (No. RS-2019-II190079, Artificial Intelligence Graduate School Program (Korea University)).}
}

\author{\IEEEauthorblockN{Sung-Jin Kim}
\IEEEauthorblockA{\textit{Dept. of Artificial Intelligence} \\
\textit{Korea University} \\
Seoul, Republic of Korea \\
s\_j\_kim@korea.ac.kr} \\
\and
\IEEEauthorblockN{Dae-Hyeok Lee}
\IEEEauthorblockA{\textit{Dept. of Brain and Cognitive Engineering} \\
\textit{Korea University} \\
Seoul, Republic of Korea \\
lee\_dh@korea.ac.kr}
\and
\IEEEauthorblockN{Hyeon-Taek Han}
\IEEEauthorblockA{\textit{Dept. of Artificial Intelligence} \\
\textit{Korea University} \\
Seoul, Republic of Korea \\
ht\_han@korea.ac.kr}
}

\maketitle

\begin{abstract}
Electroencephalography (EEG) is a generally used neuroimaging approach in brain-computer interfaces due to its non-invasive characteristics and convenience, making it an effective tool for understanding human intentions. Therefore, recent research has focused on decoding human intentions from EEG signals utilizing deep learning methods. However, since EEG signals are highly susceptible to noise during acquisition, there is a high possibility of the existence of noisy data in the dataset. Although pioneer studies have generally assumed that the dataset is well-curated, this assumption is not always met in the EEG dataset. In this paper, we addressed this issue by designing a dataset refinement algorithm that can eliminate noisy data based on metrics evaluating data influence during the training process. We applied the proposed algorithm to two motor imagery EEG public datasets and three different models to perform dataset refinement. The results indicated that retraining the model with the refined dataset consistently led to better generalization performance compared to using the original dataset. Hence, we demonstrated that removing noisy data from the training dataset alone can effectively improve the generalization performance of deep learning models in the EEG domain.
\end{abstract}


\section{INTRODUCTION}
Brain-computer interface (BCI) is a system that facilitates communication between humans and computers by interpreting human intentions \cite{han2020classification, lee2020continuous, nicolas2012brain, prabhakar2020framework, kim2015abstract}. One of the most widely utilized approaches to capture human intentions is electroencephalography (EEG). EEG is a non-invasive method for measuring brain activity, which has advantages such as portability and convenience \cite{richer2020motion}. However, due to its non-invasive approach, EEG is more susceptible to various noises compared to invasive methods \cite{nicolas2012brain}. Therefore, research has been conducted to decode human intentions embedded in EEG signals \cite{jeong2019classification}. In particular, since neural network-based deep learning has demonstrated robust performance in pattern recognition across different domains, many studies have focused on applying deep learning methodologies to decode EEG signals \cite{mane2020multi, song2022eeg}.

Neural network-based deep learning operates by learning the underlying patterns in data through the use of large amounts of training data \cite{kim2021fre}. This necessitates the availability of high-quality datasets. However, EEG signals, which are bio-signals, are prone to noise due to various factors during signal acquisition. Moreover, EEG signals exhibit significant variability in data distribution across sessions and subjects \cite{nicolas2012brain}. To address these issues, previous research has focused on refining model architectures to enhance pattern recognition capabilities \cite{kim2024towards} or applying domain adaptation techniques \cite{zhao2020deep} to reduce distributional discrepancies. Nonetheless, these methods typically assume that the training datasets are composed of finely curated data. We assumed that this strong assumption is difficult to meet in EEG datasets, which are inherently noisy and often lack quality verification. Therefore, before applying advanced methodologies to enhance performance, we assumed the refining process of the dataset was necessary. Hence, we considered the use of data pruning techniques to refine the dataset.

Generally, data pruning is a method to reduce training time in domains where the dataset is vast, and the model has numerous parameters \cite{dosovitskiy2021an}. Therefore, the primary objective of conventional data pruning is to identify easy samples that have minimal impact on training and can be removed with negligible performance degradation \cite{qin2024infobatch}. However, we hypothesized that EEG datasets, despite their limited size, contain numerous data that hinder the training of deep learning models. In this environment, a few noisy samples with high influence can significantly degrade the model's generalization performance. Hence, we aimed to improve the generalization performance of models by employing commonly used data pruning techniques, not to remove easy samples but rather to identify and eliminate hard samples.

In this paper, we propose the dataset refinement algorithm aimed at improving model generalization performance. First, we utilized various metrics to assess the influence of each data during the training process. Based on these influence assessments, we designed an algorithm to refine the dataset. To validate the effectiveness of our proposed method, we employed a motor imagery-based EEG dataset, which is particularly susceptible to noise and lacks visual confirmation of noise presence. We evaluated the robustness of our algorithm across different model architectures and datasets, using two public motor imagery datasets and three commonly used baseline models. As a result, we observed that refining the dataset using our proposed dataset refinement algorithm can improve the model's generalization performance. We demonstrate that even without employing complex methods, simply refining the dataset can enhance the model's generalization performance.

\section{METHODS}
\subsection{Metric for Measuring Influence}
We utilized two approaches to quantitatively measure the influence of data during the training process: the influence score \cite{koh2017understanding} and Monte Carlo dropout (MC dropout) \cite{deodato2019bayesian}. The influence score measures how much a particular data impacts the model's predictions. This score is widely used to analyze the effects of outliers or noisy data on the model. The following equation represents the calculation method for the influence score: 

\begin{equation}
\mathcal{I}(x_i) = \sum_{(x, y)\in\mathcal{D}}\nabla_{\theta}\mathcal{L}(x, y;\theta)^\top H_{\theta}^{-1}\nabla_{\theta}\mathcal{L}(x_i, y_i;\theta),
\end{equation}
where $\mathcal{D}$ indicates the training dataset which contains $n$ number of data and $x$ and $y$ implies data and label, respectively. $\mathcal{L}$ is the objective function. $\theta$ and $H$ inciates the emperical risk minimizer and Hessian matrix, respectively. That is, $\theta=\argmin_\theta\frac{1}{n}\sum_{i=1}^n\mathcal{L}(x_i, y_i)$ and $H_\theta=\sum_{i=1}^n\nabla_\theta^2\mathcal{L}(x_i, y_i)$.

Data with high uncertainty are more likely to be considered outliers or noisy data. To measure aleatoric uncertainty, we adopted the uncertainty quantification method proposed by Deodato \textit{et al}.\cite{deodato2019bayesian}, which is based on MC dropout. This quantification approach can be formulated as:

\begin{equation}
\mathcal{U}(x_i)=\frac{1}{T}\sum_{t=1}^Tp_{t,i}-p_{t,i}^2,
\end{equation}
where $T$ indicates the number of repetition, and $p_{t,i}$ is the confidence score of model from $x_i$ at $t^{\text{th}}$ dropout.

\subsection{Algorithm for Refining Dataset}
We designed the algorithm for dataset refinement based on the scores of data's influence in the training process. The algorithm consists of three main stages. First, we get the model weights that can minimize the empirical risk using the training dataset. Based on the empirical risk minimizer obtained in the first stage, we compute data influence scores by executing the respective algorithms to measure the influence of each data. Finally, after refining the training dataset by removing data that exhibited the highest influence score and identifying them as noisy samples, we retrain the model with the refined dataset. Through this iterative process, we successfully enhanced the model's generalization performance by eliminating noisy samples from the training dataset. The details of the proposed algorithm are presented as Alg. 1.

%%%%%%%%%%%%%%%%%%%%%%%%%%%%%%%%%%%%%%%%%%%%%%%%%%%%%%%%%%%%%%%%%%%%%%%%%%%%%%%%%%%%%%%%%%%%%%%%%%%%

\begin{algorithm}[t]
\caption{The Proposed Algorithm}
\begin{algorithmic}
\STATE 
\textbf{Input:} Training dataset $\mathcal{D}=\{(x_i, y_i)\}^n_{i=1}$, learning rate $\lambda$ \\
cross-entropy function $H$, the number of training epochs $E$, \\
cardinality of dataset $|\cdot|$ \\
\textbf{Output:} empirical risk minimizer $\theta'$ with the refined dataset

\STATE {\textbf{Stage I}: Train the model $f$ utilizing $\mathcal{D}$} 
\STATE \hspace{0.25cm} \textbf{for} epoch $= 1$ \textbf{to} $E$ \textbf{do}
\STATE \hspace{0.75cm} $ \mathcal{L} = -\frac{1}{n}\sum_{i=1}^{n}H(y_i, f(x_i;\theta)) $
\STATE \hspace{0.75cm} $ \theta \leftarrow \theta - \lambda \nabla_{\theta}\mathcal{L} $
\STATE \hspace{0.25cm} \textbf{end for}

\STATE {\textbf{Stage II}: Calculate the influence of data}
\STATE \hspace{0.25cm} \textbf{for each} data $(x_i, y_i)$ in $\mathcal{D}$ \textbf{do}
\STATE \hspace{0.75cm} \textbf{if} metric is influence score \textbf{do}
\STATE \hspace{1.25cm} $ s_i = \mathcal{I}(x_i) $
\STATE \hspace{0.75cm} \textbf{else if} metric is MC dropout \textbf{do}
\STATE \hspace{1.25cm} $ s_i = \mathcal{U}(x_i) $
\STATE \hspace{0.75cm} \textbf{end if}
\STATE \hspace{0.25cm} \textbf{end for}
\STATE \hspace{0.25cm} \textbf{return}  $\{s_1, s_2, \dots, s_n\}$

\STATE {\textbf{Stage III}: Retrain the model with eliminating noisy samples}
\STATE \hspace{0.40cm}1. Optimize the threshold $\tau$ for identifying noisy samples
\STATE \hspace{0.40cm}2. Remove the data considered as noisy samples
\STATE \hspace{0.80cm}$ \mathcal{D}_\text{noise} = \{x_i \in \mathcal{D} \mid s_i < \tau, i \in [1, n] \}$
\STATE \hspace{0.40cm}3. Get the empirical risk minimizer $\theta'$ from $\mathcal{D}\backslash\mathcal{D}_\text{noise}$
\STATE \hspace{0.70cm} $ m = |\mathcal{D}| - |\mathcal{D_{\text{noise}}}| $
\STATE \hspace{0.70cm} \textbf{for} epoch $= 1$ \textbf{to} $E$ \textbf{do}
\STATE \hspace{1.10cm} $ \mathcal{L} = -\frac{1}{m}\sum_{i=1}^{m}H(y_i, f(x_i;\hat{\theta})) $
\STATE \hspace{1.10cm} $ \hat{\theta} \leftarrow \hat{\theta} - \lambda \nabla_{\hat{\theta}}\mathcal{L} $
\STATE \hspace{0.70cm} \textbf{end for}

\end{algorithmic}
\end{algorithm}

%%%%%%%%%%%%%%%%%%%%%%%%%%%%%%%%%%%%%%%%%%%%%%%%%%%%%%%%%%%%%%%%%%%%%%%%%%%%%%%%%%%%%%%%%%%%%%%%%%%%

\subsection{Datasets}
To evaluate the proposed algorithm, we utilized two public motor imagery-based EEG datasets that are most generally used for assessment in the motor imagery EEG domain: BCI Competition IV 2a \cite{brunner2008bci} and BCI Competition IV 2b \cite{leeb2008bci}.

BCI Competition IV 2a (BCIC2a) was acquired from nine subjects performing motor imagery tasks using 22 electrodes placed according to the international 10/20 system. The electrodes were set with a sampling rate of 250 Hz. The subjects were instructed to perform four different motor imagery tasks: imagining movements of the right hand, left hand, tongue, and foot. EEG signals were collected over two sessions per subject, each consisting of 288 trials, corresponding to 72 repetitions per class. We defined the imagery segment as the period between 2 to 6 seconds and applied a low-pass Butterworth filter at 38 Hz to focus on the target frequency range associated with motor imagery.

%%%%%%%%%%%%%%%%%%%%%%%%%%%%%%%%%%%%%%%%%%%%%%%%%%%%%%%%%%%%%%%%%%%%%%%%%%%%%%%%%%%%%%%%%%%%%%%%%%%%
\begin{figure*}[t!]
\centering
\scriptsize
\centerline{\includegraphics[width=0.9\textwidth]{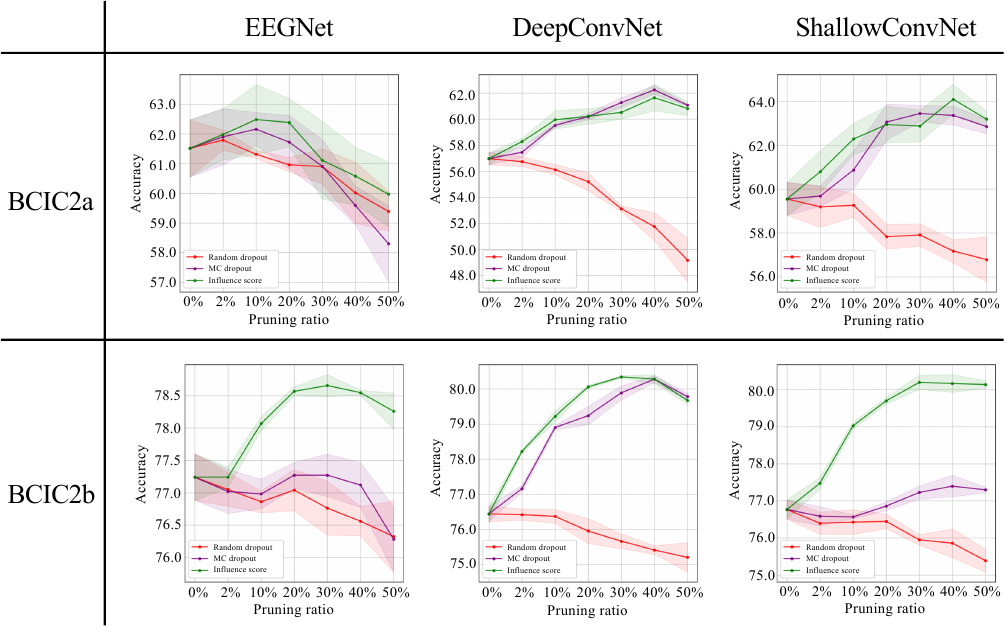}}
\caption{Comparison of performances based on the ratio of dataset refinement in two different datasets and three models. The shade indicates the standard deviation calculated from ten different random seeds.}
\end{figure*}
%%%%%%%%%%%%%%%%%%%%%%%%%%%%%%%%%%%%%%%%%%%%%%%%%%%%%%%%%%%%%%%%%%%%%%%%%%%%%%%%%%%%%%%%%%%%%%%%%%%%

BCI Competition IV 2b (BCIC2b) was collected from nine subjects who performed right-hand and left-hand motor imagery tasks while three electrodes were placed over the sensory-motor cortex. The electrodes were configured with the sampling rate of 250 Hz, similar to BCIC2a. Data was collected over five sessions for each subject. In the first two sessions, 120 trials were conducted similarly to BCIC2a. In the remaining three sessions, subjects received visual feedback about their imagery results, with 160 trials per session. The target imagery segment was set from 3 to 7 seconds, and a low-pass Butterworth filter at 38 Hz was applied to focus on the target frequency.

\subsection{Evaluation Settings}
To evaluate the robustness and effectiveness of our proposed algorithm, we applied it to three different models generally used in the motor imagery domain. The selected models were EEGNet \cite{lawhern2018eegnet}, DeepConvNet \cite{schirrmeister2017deep}, and ShallowConvNet \cite{schirrmeister2017deep, kim2022rethinking}. Each model was trained from scratch, with a learning rate of 2e-3 for 300 epochs and ten epochs warm-up phase to stabilize training. These hyperparameters were consistently applied across both datasets. We used a cosine learning rate scheduler and the AdamW optimizer to stabilize the training process. The batch size was set to 64. For evaluation, we employed leave-one-subject-out cross-validation to conduct experiments in a subject-independent setting. The evaluation metric was average accuracy, which is widely used in subject-independent motor imagery domains \cite{dornhege2007toward, ahn2022multiscale}, and we reported results obtained over ten different runs with various random seeds. To verify whether the performance gain of the proposed algorithm was solely due to data reduction, we additionally introduced a random dropout approach that randomly removes data. The threshold for distinguishing noisy data was determined through the grid search, and all corresponding results were reported. For data preprocessing, we used exponential moving standardization to remove outliers within the signal \cite{zhao2020deep}, and the respective formulation is provided below.

\begin{align} 
& x'_{i, k}=\frac{x_{i, k}-\mu_k}{\sqrt{\sigma_k^2}}, \\
& \mu_k=(1-\alpha)x_{i, k}+\alpha\mu_{k-1}, \\
& \sigma_k^2=(1-\alpha)(x_{i, k}-\mu_k)^2+\alpha\sigma_{k-1}^2,    
\end{align}
where $k$ is the index of time point in data $x_i$, and $x'_{i, k}$ indicates the preprocessed data. We set the $\alpha$ as 0.999.

\section{RESULTS AND DISCUSSION}
As shown in Fig. 1, we evaluated the performance of the proposed algorithm by applying it to three different models on two different datasets. The results indicate that refining the dataset using influence scores and MC dropout consistently led to performance improvement compared to using the original dataset. For the BCIC2a dataset, the highest performance improvement was observed when refining 10 \% of the dataset for EEGNet and 40 \% for both DeepConvNet and ShallowConvNet, resulting in performance gains of 0.97 \%, 4.67 \%, and 4.54 \%, respectively. For the BCIC2b dataset, the highest improvements were obtained by refining 30 \% of the data, with EEGNet achieving 1.42 \%, DeepConvNet achieving 3.90 \%, and ShallowConvNet achieving 3.43 \% gains. For MC dropout, the BCIC2a dataset showed performance improvements of 0.65 \%, 5.27 \%, and 3.89 \% for EEGNet, DeepConvNet, and ShallowConvNet, respectively, when refining 10 \%, 40 \%, and 30 \% of the dataset, respectively. For the BCIC2b dataset, performance gains of 0.03 \%, 3.11 \%, and 1.81 \% were observed for EEGNet, DeepConvNet, and ShallowConvNet, respectively, when refining 20 \%, 40 \%, and 40 \% of the dataset, respectively. The influence score provided more significant improvements in model generalization performance compared to MC dropout.

We also observed that the proposed algorithm was particularly effective in models with a relatively larger number of parameters, such as DeepConvNet and ShallowConvNet. This suggests the potential for further performance improvements when applying the proposed algorithm to more complex models. Additionally, we verified that dataset refinement using the proposed algorithm yielded more significant performance improvements compared to random dropout. Hence, we demonstrate that our approach effectively enhances model generalization performance.

\section{CONCLUSION AND FUTURE WORKS}
In this paper, we proposed the algorithm to refine EEG datasets to improve decoding models' generalization performance. Pioneer studies on extracting significant features from EEG signals have been conducted in various directions, such as enhancing model architectures and applying domain adaptation. However, these methods often rely on the assumption that the dataset is well-curated, which is difficult to achieve with EEG data due to its susceptibility to noise. To address this issue, we proposed the algorithm that applies various metrics to quantify data influence during training, allowing for effective dataset refinement. Our results demonstrated that refining the dataset using the proposed algorithm resulted in significant improvements in model performance, with up to 5.27 \% improvement on the BCIC2a dataset and up to 3.90 \% improvement on the BCIC2b dataset. This improvement was achieved simply by eliminating data that adversely affected the training process, thus enhancing model generalization.

Furthermore, by validating the robustness of the proposed algorithm by applying it to three different models across two different datasets, we observed consistent performance improvements. The contribution of the proposed algorithm lies in its applicability as a preprocessing technique alongside other existing methods. Despite the effectiveness of the algorithm in removing noisy data from the dataset, there remain limitations. Specifically, the process of calculating influence scores and optimizing the threshold requires additional computational resources and costs. While this approach was feasible for relatively small datasets like BCIC2a and BCIC2b, applying it to larger datasets may present challenges. Therefore, our future work will focus on developing a more efficient dataset refinement algorithm that can effectively operate even with large-scale datasets.

\bibliographystyle{IEEEtran}
\bibliography{REFERENCE}

\begin{thebibliography}{10}
\providecommand{\url}[1]{#1}
\csname url@samestyle\endcsname
\providecommand{\newblock}{\relax}
\providecommand{\bibinfo}[2]{#2}
\providecommand{\BIBentrySTDinterwordspacing}{\spaceskip=0pt\relax}
\providecommand{\BIBentryALTinterwordstretchfactor}{4}
\providecommand{\BIBentryALTinterwordspacing}{\spaceskip=\fontdimen2\font plus
\BIBentryALTinterwordstretchfactor\fontdimen3\font minus \fontdimen4\font\relax}
\providecommand{\BIBforeignlanguage}[2]{{%
\expandafter\ifx\csname l@#1\endcsname\relax
\typeout{** WARNING: IEEEtran.bst: No hyphenation pattern has been}%
\typeout{** loaded for the language `#1'. Using the pattern for}%
\typeout{** the default language instead.}%
\else
\language=\csname l@#1\endcsname
\fi
#2}}
\providecommand{\BIBdecl}{\relax}
\BIBdecl

\bibitem{han2020classification}
S.-Y. Han, N.-S. Kwak, T.~Oh, and S.-W. Lee, ``{Classification of pilots’ mental states using a multimodal deep learning network},'' \emph{Biocybern. Biomed. Eng.}, vol.~40, no.~1, pp. 324--336, 2020.

\bibitem{lee2020continuous}
D.-H. Lee, J.-H. Jeong, K.~Kim, B.-W. Yu, and S.-W. Lee, ``{Continuous EEG decoding of pilots’ mental states using multiple feature block-based convolutional neural network},'' \emph{IEEE Access}, vol.~8, pp. 121\,929--121\,941, 2020.

\bibitem{nicolas2012brain}
L.~F. Nicolas-Alonso and J.~Gomez-Gil, ``{Brain computer interfaces, a review},'' \emph{Sensors}, vol.~12, no.~2, pp. 1211--1279, 2012.

\bibitem{prabhakar2020framework}
S.~K. Prabhakar, H.~Rajaguru, and S.-W. Lee, ``{A framework for schizophrenia EEG signal classification with nature inspired optimization algorithms},'' \emph{IEEE Access}, vol.~8, pp. 39\,875--39\,897, 2020.

\bibitem{kim2015abstract}
J.~Kim \emph{et~al.}, ``{Abstract representations of associated emotions in the human brain},'' \emph{J. Neurosci. Res.}, vol.~35, no.~14, pp. 5655--5663, 2015.

\bibitem{richer2020motion}
N.~Richer, R.~J. Downey, W.~D. Hairston, D.~P. Ferris, and A.~D. Nordin, ``{Motion and muscle artifact removal validation using an electrical head phantom, robotic motion platform, and dual layer mobile EEG},'' \emph{IEEE Trans. Neural Syst. Rehabil. Eng.}, vol.~28, no.~8, pp. 1825--1835, 2020.

\bibitem{jeong2019classification}
J.-H. Jeong, B.-W. Yu, D.-H. Lee, and S.-W. Lee, ``{Classification of drowsiness levels based on a deep spatio--temporal convolutional bidirectional LSTM network using electroencephalography signals},'' \emph{Brain Sci.}, vol.~9, no.~12, p. 348, 2019.

\bibitem{mane2020multi}
R.~Mane, N.~Robinson, A.~P. Vinod, S.-W. Lee, and C.~Guan, ``{A multi--view CNN with novel variance layer for motor imagery brain computer interface},'' in \emph{Proc. Int. Conf. IEEE Eng. Med. Biol. Soc. (EMBC)}, 2020, pp. 2950--2953.

\bibitem{song2022eeg}
Y.~Song, Q.~Zheng, B.~Liu, and X.~Gao, ``{EEG conformer: Convolutional transformer for EEG decoding and visualization},'' \emph{IEEE Trans. Neural Syst. Rehabil. Eng.}, vol.~31, pp. 710--719, 2022.

\bibitem{kim2021fre}
J.-H. Kim, S.-H. Lee, J.-H. Lee, and S.-W. Lee, ``{Fre-GAN: Adversarial frequency-consistent audio synthesis},'' in \emph{Proc. INTERSPEECH}, 2021, pp. 3246--3250.

\bibitem{kim2024towards}
S.-J. Kim, D.-H. Lee, H.-G. Kwak, and S.-W. Lee, ``{Towards domain-free transformer for generalized EEG pre-training},'' \emph{IEEE Trans. Neural Syst. Rehabil. Eng.}, 2024.

\bibitem{zhao2020deep}
H.~Zhao, Q.~Zheng, K.~Ma, H.~Li, and Y.~Zheng, ``{Deep representation--based domain adaptation for nonstationary EEG classification},'' \emph{IEEE Trans. Neural Netw. Learn. Syst.}, vol.~32, no.~2, pp. 535--545, 2020.

\bibitem{dosovitskiy2021an}
A.~Dosovitskiy \emph{et~al.}, ``{An image is worth 16x16 words: Transformers for image recognition at scale},'' in \emph{Proc. Int. Conf. Learn. Represent. (ICLR)}, 2021.

\bibitem{qin2024infobatch}
Z.~Qin \emph{et~al.}, ``{InfoBatch: Lossless training speed up by unbiased dynamic data pruning},'' in \emph{Proc. Int. Conf. Learn. Represent. (ICLR)}, 2024.

\bibitem{koh2017understanding}
P.~W. Koh and P.~Liang, ``{Understanding black--box predictions via influence functions},'' in \emph{Proc. Int. Conf. Mach. Learn. (ICML)}, 2017, pp. 1885--1894.

\bibitem{deodato2019bayesian}
G.~Deodato, C.~Ball, and X.~Zhang, ``{Bayesian neural networks for cellular image classification and uncertainty analysis},'' \emph{bioRxiv}, p. 824862, 2019.

\bibitem{brunner2008bci}
C.~Brunner, R.~Leeb, G.~M{\"u}ller-Putz, A.~Schl{\"o}gl, and G.~Pfurtscheller, ``{BCI competition 2008-Graz data set A},'' \emph{Inst. Knowl. Discovery, Lab. Brain-Comput. Interfaces, Graz Univ. Technol.}, vol.~16, pp. 1--6, 2008.

\bibitem{leeb2008bci}
R.~Leeb, C.~Brunner, G.~M{\"u}ller-Putz, A.~Schl{\"o}gl, and G.~Pfurtscheller, ``{BCI competition 2008-Graz data set B},'' \emph{Graz Univ. Technol., Austria}, pp. 1--6, 2008.

\bibitem{lawhern2018eegnet}
V.~J. Lawhern \emph{et~al.}, ``{EEGNet: A compact convolutional neural network for EEG-based brain-computer interfaces},'' \emph{J. Neural Eng.}, vol.~15, no.~5, p. 056013, 2018.

\bibitem{schirrmeister2017deep}
R.~T. Schirrmeister \emph{et~al.}, ``{Deep learning with convolutional neural networks for EEG decoding and visualization},'' \emph{Hum. Brain Mapp.}, vol.~38, no.~11, pp. 5391--5420, 2017.

\bibitem{kim2022rethinking}
S.-J. Kim, D.-H. Lee, and S.-W. Lee, ``{Rethinking CNN architecture for enhancing decoding performance of motor imagery-based EEG signals},'' \emph{IEEE Access}, vol.~10, pp. 96\,984--96\,996, 2022.

\bibitem{dornhege2007toward}
G.~Dornhege, J.~R. Mill{\'a}n, T.~Hinterberger, D.~J. McFarland, and K.-R. M{\"u}ller, \emph{{Evaluation criteria for BCI research}}.\hskip 1em plus 0.5em minus 0.4em\relax MIT Press, 2007, ch. {Toward Brain--Computer Interfacing}, pp. 327--342.

\bibitem{ahn2022multiscale}
H.-J. Ahn, D.-H. Lee, J.-H. Jeong, and S.-W. Lee, ``{Multiscale convolutional transformer for EEG classification of mental imagery in different modalities},'' \emph{IEEE Trans. Neural Syst. Rehabil. Eng.}, vol.~31, pp. 646--656, 2022.

\end{thebibliography}

\end{document}